\documentclass[twocolumn,12pt]{article}

\usepackage{arxiv}

\usepackage[utf8]{inputenc} 
\usepackage[T1]{fontenc}    
\usepackage{hyperref}       
\usepackage{url}            
\usepackage{booktabs}       
\usepackage{amsfonts}       
\usepackage{nicefrac}       
\usepackage{microtype}      
\usepackage{lipsum}		
\usepackage{graphicx}
\usepackage[numbers]{natbib}
\usepackage{doi}
\usepackage{abstract}

\usepackage{geometry}
\renewenvironment{abstract}
{
  \centerline
  {\large \bfseries \scshape Abstract}
  \begin{quote}
  \normalsize 
}
{
  \end{quote}
}

\title{The Protein Engineering Tournament: \\ An Open Science Benchmark for Protein Modeling and Design}

\author{
    Chase Armer\textsuperscript{1*}\href{https://orcid.org/0000-0003-0951-3198}{\includegraphics[scale=0.06]{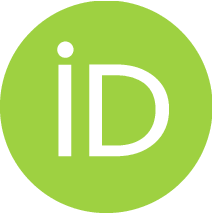}},
    Hassan Kane\textsuperscript{2*}\href{https://orcid.org/0009-0005-6937-6385}{\includegraphics[scale=0.06]{orcid.eps}},
    Dana Cortade\textsuperscript{3*}\href{https://orcid.org/0000-0002-9267-4015}{\includegraphics[scale=0.06]{orcid.eps}},
    Dave Estell\textsuperscript{4},
    Adil Yusuf\textsuperscript{2}\href{https://orcid.org/0009-0003-8496-1908}{\includegraphics[scale=0.06]{orcid.eps}},
    Radhakrishna Sanka\textsuperscript{5},\\
    \textbf{
    Henning Redestig\textsuperscript{4}\href{https://orcid.org/0000-0003-2130-9288}{\includegraphics[scale=0.06]{orcid.eps}},
    TJ Brunette\textsuperscript{3}\href{https://orcid.org/0000-0003-0748-8224}{\includegraphics[scale=0.06]{orcid.eps}},
    Pete Kelly\textsuperscript{3},
    Erika DeBenedictis\textsuperscript{3,6}\href{https://orcid.org/0000-0002-7933-2651}{\includegraphics[scale=0.06]{orcid.eps}}
    }
}

\date{
\small{
    \textsuperscript{1}Columbia University, New York City, United States \ \hspace{0.3em} \textbullet\ \hspace{0.3em} \textsuperscript{2}Medium Biosciences \\
    \textsuperscript{3}Align to Innovate, Cambridge, United States \ \hspace{0.3em} \textbullet\ \hspace{0.3em} \textsuperscript{4}International Flavors and Fragrances \\
    \textsuperscript{5}Boston University, Boston, United States \ \hspace{0.3em} \textbullet\ \hspace{0.3em} \textsuperscript{6}Francis Crick Institute, London, United Kingdom \\
    \textsuperscript{\textasteriskcentered}Designates equal contribution \\
    \href{mailto:tournament@alignbio.org}{E-mail: tournament@alignbio.org}
}}

\renewcommand{\undertitle}{}
\renewcommand{\shorttitle}{The Protein Engineering Tournament}

\hypersetup{
pdftitle={The Protein Engineering Tournament: \\ An Open Science Benchmark for Protein Modeling and Design},
pdfsubject={q-bio.QM},
pdfauthor={Chase Armer},
pdfkeywords={Protein Engineering, Computation Modeling, Open Source, Benchmarking, Cloud Laboratories},
}

\begin{document}

\twocolumn[
\begin{@twocolumnfalse}
    \maketitle
    \begin{abstract}
The grand challenge of protein engineering is the development of computational models that can characterize and generate protein sequences for any arbitrary function. However, progress today is limited by lack of 1) benchmarks with which to compare computational techniques, 2) large datasets of protein function, and 3) democratized access to experimental protein characterization. Here, we introduce the Protein Engineering Tournament, a fully-remote, biennial competition for the development and benchmarking of computational methods in protein engineering. The tournament consists of two rounds: a first in silico round, where participants use computational models to predict biophysical properties for a set of protein sequences, and a second in vitro round, where participants are challenged to design new protein sequences, which are experimentally measured with open-source, automated methods to determine a winner. At the Tournament's conclusion, the experimental protocols and all collected data will be open-sourced for continued benchmarking and advancement of computational models. We hope the Protein Engineering Tournament will provide a transparent platform with which to evaluate progress in this field and mobilize the scientific community to conquer the grand challenge of computational protein engineering.
    \end{abstract}
    \vspace{1cm} 
\end{@twocolumnfalse}
] 

\begin{figure*}
    \centering
    \includegraphics[width=0.8\textwidth]{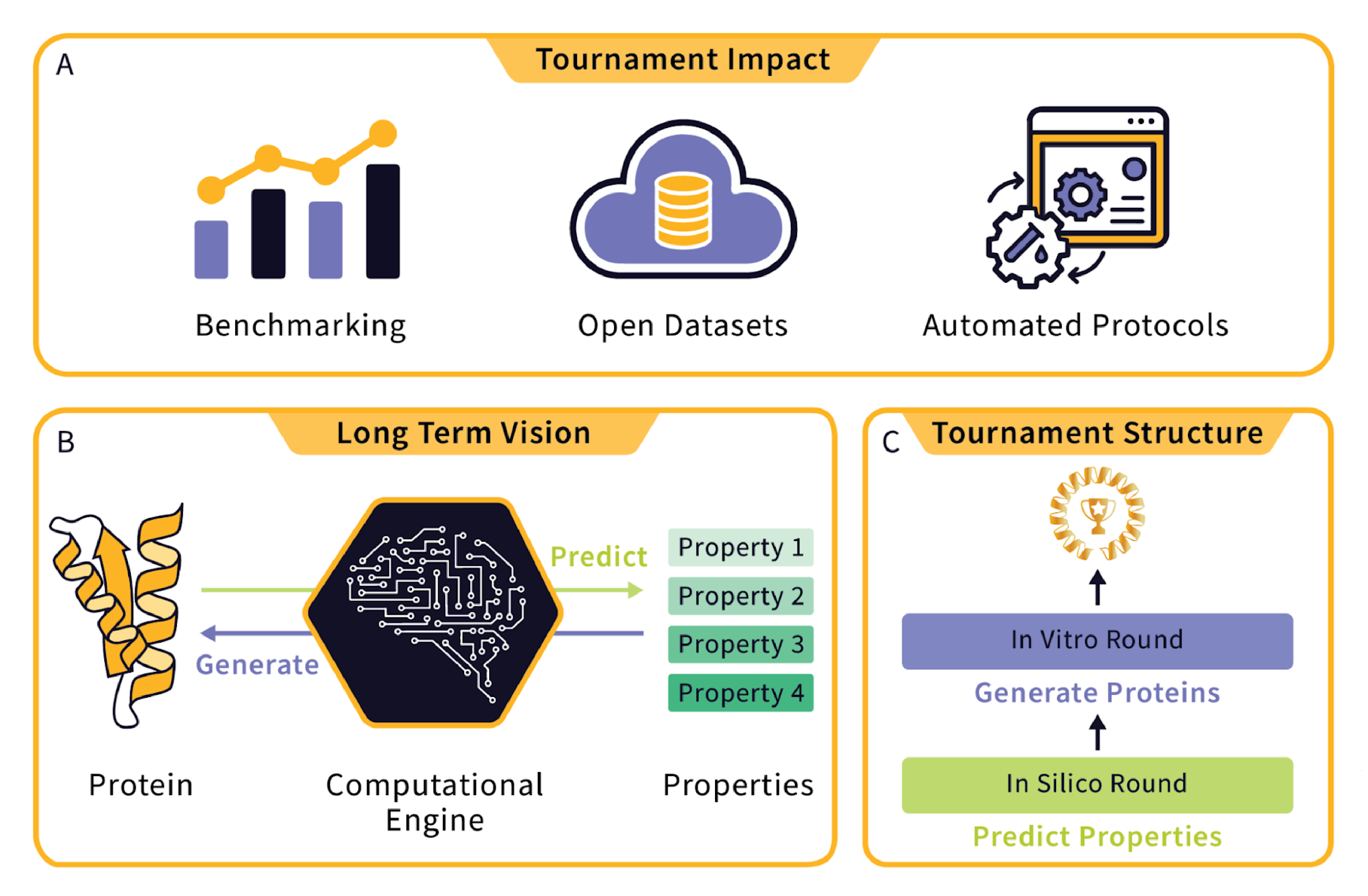}
    \caption{
        \textbf{Overview of the Tournament.} 
        \textbf{(A)}The tournament impacts the space by providing a transparent benchmark for computational methods, open datasets for the research community, and automated protocols for continued independent benchmarking. 
        \textbf{(B)} The Tournament is designed to accelerate creation of computational techniques that can predict the biophysical properties for any given protein and generate a protein with any desired properties. 
        \textbf{(C)} The Tournament consists of two sequential rounds: an in silico round for predicting properties of proteins and an in vitro round for generating proteins with specific properties.
    }
    \label{fig:figure1}
\end{figure*}

\section{Introduction}
Protein engineering is the discipline of modifying or designing protein DNA sequences to create new functions or improve upon existing ones\citep{ulmer1983protein}. It has historically been used to create new enzymes for industrial processes\citep{bornscheuer2012engineering}, develop new biologics with improved efficacy\citep{maynard2000antibody}, and engineer proteins for better tasting plant-based burgers\citep{whitehurst2010enzymes}. Many current global challenges are also being addressed through protein engineering, such as: reducing greenhouse gas emissions\citep{fradette2017enzyme}, recycling plastic\citep{gao2021recent}, diagnosing biomarkers related to rare diseases\citep{singh2019enzymes} and the development of biologics targeting infectious diseases\citep{fan2017engineering}. In order to develop increasingly sophisticated protein solutions, powerful computational models are needed to guide the design of proteins with new or improved functionality.
\newpage

The field of computational protein engineering aims to guide the design of proteins by developing predictive\citep{hie2022adaptive} and generative\citep{strokach2022deep} models; however, several obstacles are currently limiting model development. Although various machine learning methods have been developed over the years to take advantage of existing evolutionary, structural, and assay data, the lack of complex, publicly available datasets, limited experimental reproducibility, and absence of infrastructure for benchmarking models all impede model development and validation. The majority of publicly available datasets\citep{wang2019protabank} capturing sequence to function data are relatively simple, mapping single point mutations\citep{notin2022tranception} to a single biophysical property per experimental condition. When designing new variants, computational scientists often lack the means\citep{thomas2022tuned} to experimentally reproduce the conditions in which certain datasets were created and therefore cannot reliably perform comparisons against existing models. Until now, there have been no common benchmarks nor infrastructure for experimentally assessing such models.

The Protein Engineering Tournament aims to tackle the aforementioned obstacles by curating a series of tournaments centered on various protein engineering challenges. By providing a transparent platform for benchmarking protein design methods, generating publicly available datasets, and developing open-sourced infrastructure for automated experimentation, the Tournament hopes to reduce barriers to model development and validation (Figure 1A). The Tournament will act as a platform for accelerating development in computational protein design (Figure 1B), enabling computational scientists to benchmark their models by both predicting protein properties and generating novel proteins with improved functionality. To achieve this goal, the Tournament includes both an in silico round and an in vitro round (Figure 1C). 

The Tournament also aims to inspire more machine learning scientists to contribute to the field of protein engineering by making the field more accessible and transparent. The Tournament will engage with the field of machine learning to solve technical problems that uniquely arise in protein engineering and develop the infrastructure it needs to design, build, and test better models and protein engineering strategies.  Furthermore, we hope to engage with the community by showcasing challenges with high societal value that are currently unsolved. To achieve this goal, we will intentionally focus on both technically challenging tasks and use cases which may be overlooked by academia and/or industry.

\begin{figure*}
    \centering
    \includegraphics[width=0.8\textwidth]{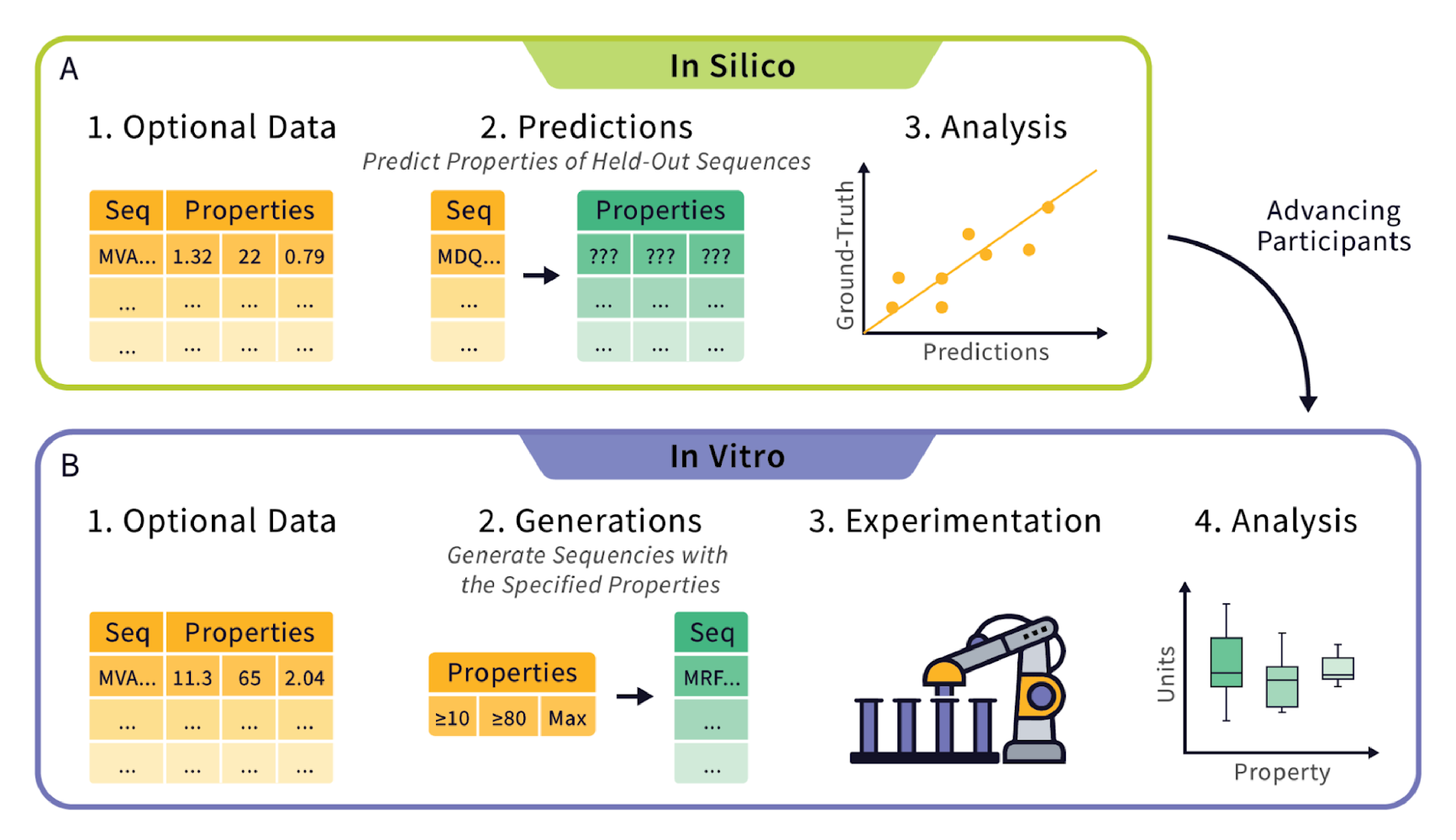}
    \caption{
        \textbf{Overview of Tournament Rounds.} The tournament consists of two rounds. 
        \textbf{(A)} In the in silico round participants will predict properties of given protein sequences, with optional training data provided for specific events. Performance is evaluated by comparing the participant’s predictions with the ground-truth values. Top performing participants will be selected to advance to the in vitro round. 
        \textbf{(B)} In the in vitro round participants will be asked to generate protein sequences based on desired properties, which will then be expressed and characterized using cloud labs. Performance of the designed sequences will be evaluated as a weighted combination of the protein’s properties; the exact evaluation metric will be event-dependent. 
    }
    \label{fig:figure2}
\end{figure*}

\section{Related Approaches}
Open datasets have long provided valuable opportunities for developing and benchmarking new methods in machine learning research. Computer vision datasets, such as MNIST\citep{deng2012mnist} and ImageNet\citep{deng2009imagenet}, not only provided individual research labs with a substrate for experimenting with new approaches but also created a yardstick with which to measure collective progress. Researchers in the protein engineering community have also utilized open datasets and introduced tasks, such as FLIP\citep{dallago2021flip}, TAPE\citep{rao2019evaluating}, and ProteinNet\citep{alquraishi2019proteinnet}, to encourage similar developments. These datasets have challenged researchers to develop models performing a wide range of tasks, such as predicting biophysical properties on both protein and amino acid levels, as well as predictions on variant effects and structural features. 

Science competitions take these efforts a step further by allowing researchers to test their computational methods on never-before-seen datasets. Perhaps the most notable example is the Critical Assessment of Structure Prediction (CASP)\citep{moult1995large} a biennial event for computational protein structure prediction. Since its inception, CASP has become a crucial benchmark for the protein structure prediction community. By creating visibility around a single event, the competition has inspired an ambitious spirit among researchers to develop the best performing method, thereby encouraging a strong pace of 

method development. Moreover, a well-known competition can offer an accessible entry point for research groups outside of the field to participate, as demonstrated by Deepmind’s participation in the 2018 CASP competition\citep{moult2018critical}. CASP has inspired the creation of similar competitions, like the Critical Assessment of Computational Hit-finding Experiments (CACHE)\citep{ackloo2022cache}, which was created in the computational chemistry field to benchmark novel approaches for finding new small-molecule binders.

We believe there is a burgeoning opportunity to create a new scientific competition that addresses the unique challenges of predicting and engineering protein function. Computational research groups which lack the ability to experimentally characterize engineered proteins are currently unable to meaningfully evaluate the performance of their protein engineering methods. By introducing never-before-seen datasets on protein function and offering open-source experimental characterization of novel proteins, the Protein Engineering Tournament hopes to overcome this barrier and enable research groups from all backgrounds to participate in cutting-edge protein engineering research. In doing so, we expect the Tournament will become a unifying benchmark for the field.

\section{THE PROTEIN ENGINEERING TOURNAMENT}
\subsection{Tournament structure}
The tournament will consist of two sequential rounds: the in silico round and the in vitro round (Figure 2).  

In the in silico round teams will be tasked with developing predictive models which can infer the biophysical properties of protein sequences (Figure 2A). The in silico round contains the option to either directly predict biophysical properties based on protein sequence (zero-shot learning) or to pre-train a model on an optional training dataset (supervised learning). After choosing their method, participants will be asked to predict the biophysical properties for a held-out test set of protein sequences, such as their stability, expressibility, and activity. Submissions will be evaluated by a comparison between predictions and experimental data, using statistics such as the Spearman correlation. 

Each biophysical property will be assessed independently, and participants will be allowed to submit predictions for as few or as many of these properties as they desire. Therefore, research labs that are building specialized computational tools for one specific property of interest, such as a stability predictor or an enzymatic activity predictor, will be able to focus on submitting predictions for that property alone. In contrast, research teams focused on applying state-of-the-art machine learning techniques to biological datasets, for example, may be more interested in submitting predictions for all available properties. Once submissions are closed and the performance of each team has been evaluated, the final leaderboard of the in silico round will be published and the datasets used will be made publicly available. The highest performing teams will then advance to the in vitro round. 

In the in vitro round (Figure 2B), the teams will be asked to design protein sequences that maximize or satisfy certain biophysical properties (e.g., an enzyme design challenge may ask for sequences which maximize enzymatic activity while staying above a specified threshold for stability and expressibility). Each team will submit a list of amino acid sequences that will then be synthesized and experimentally characterized using automated laboratory protocols developed by the Tournament and its partners. Once experimentally characterized, a ranking algorithm will evaluate the submissions to produce a score for each participating team. As an example, a submission score may be produced by calculating the normalized discounted cumulative gain (NDCG) for the enzymatic activity of all sequences above the required thresholds of stability and expressibility. The exact evaluation metric will depend on the protein target in question and will be tailored to the academic or industrial use-case for which it is being studied. 

At the conclusion of the in vitro round, the Tournament will publish the final leaderboard, and the team with the highest performing proteins will be awarded the title of Protein Engineering Champion. The characterized protein sequences in the in vitro round and the datasets of the  in silico round will be made publicly available. Furthermore, the automated protocols used to experimentally characterize the designed proteins will be made available to the public for continued use.

\begin{figure}
    \centering
    \includegraphics[width=1\columnwidth]{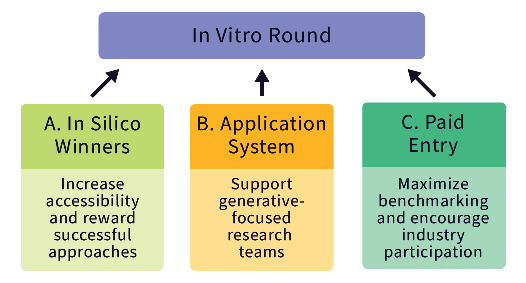}
    \caption{
        \textbf{Round Advancement.} There will be three avenues for participants to enter into the in vitro round, with each avenue catering to a unique audience.
        \textbf{(A)} Research groups specializing in generative design methods can apply to directly enter the in vitro round via an application submission.
        \textbf{(B)} The top performing teams from the in silico round will be invited to participate in the in vitro round.
        \textbf{(C)} Groups can pay to enter the in vitro round directly.}

    \label{fig:figure3}
\end{figure}

\begin{figure*}
    \centering
    \includegraphics[width=0.8\textwidth]{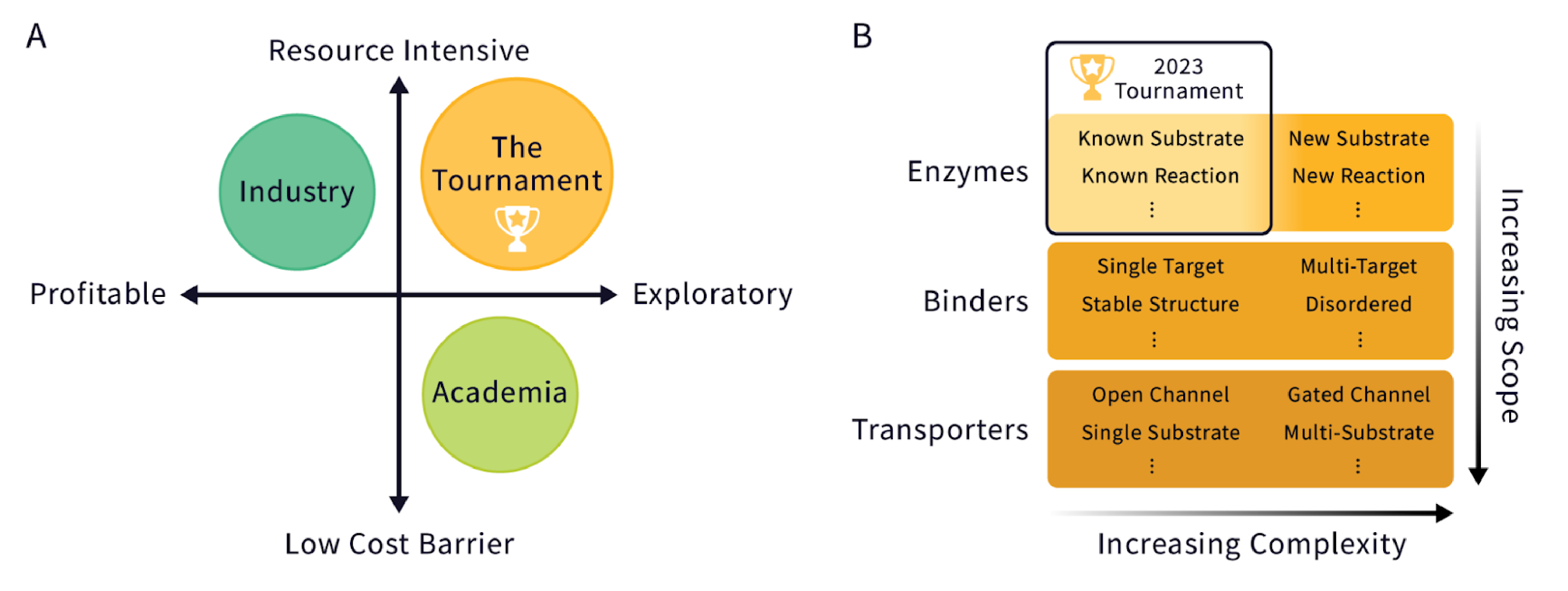}
    \caption{
        \textbf{Selecting Protein Design Challenges.} 
        \textbf{(A)} The Tournament provides a unique opportunity to highlight important protein engineering challenges that would otherwise fall outside the purview of both academic and industry incentives.
        \textbf{(B)} The Tournament will continually expand to new domains of protein design; in each domain, we will continually select challenges that push the limits of current techniques.
    }
    \label{fig:figure4}
\end{figure*}

\subsection{Participation}
The first round of the Protein Engineering Tournament, the in silico round, will be open to any and all researchers interested in participating. Interested teams composed of one or more individuals will be able to download challenge data and upload final predictions. The in silico challenge will be open to the community for a specified amount of time, after which the submissions will be evaluated and the final leaderboard will be released.  

To allow the greatest number of teams to participate in the in vitro round, while being conscious of the costs associated with DNA synthesis and experimental characterization, we propose three avenues for admission: 1) top performance in the in silico round, 2) submitting a written application, and 3) paid entry for corporate researchers (Figure 3). By offering these three avenues for participating in the in vitro challenge we will increase the accessibility of the tournament while maximizing opportunities for participation, community impact, and benchmarking of new methods.

First, the top-performing teams in the in silico round will be offered the opportunity to advance to the in vitro round (Figure 3A). Since the in silico round is open to any team or individual with access to a computer, with no entry fee or requirement on prior research experience, this path focuses on rewarding successful computational approaches and increasing the accessibility of the tournament

Second, we will provide an application system that allows researchers to apply directly for a spot in the in vitro round (Figure 3B). The value of providing an application-based path to entry is born from the fundamental differences that exist between the in silico and in vitro challenges. There are likely research groups who have developed impressive methods for generative protein design, which would be well-suited for the in vitro challenge, but have no prior experience with the property prediction tasks requested in the in silico challenge. Without a separate application process, these groups may perform poorly in the in silico round, or simply may not choose to participate at all, and their generative methods would not have the opportunity to be benchmarked in the Tournament. 

For the third and final path to entry, participants will be allowed to pay a fee which covers the cost of experimentation (Figure 3C). With this option, funding for the Protein Engineering Tournament no longer becomes the limiting factor on the number of participants that can compete and the number of methods that can be evaluated. This path could provide entry to well-funded corporate research labs capable of managing these costs, which further allows our application system discussed above to focus on providing entry to promising, less well-funded research groups. 

Our belief is that by offering these three avenues for participating in the in vitro challenge we will increase the accessibility of the tournament and maximize opportunities for participation and benchmarking of new methods.

\subsection{Selecting our protein engineering challenges}
More than benchmarking the field, we believe the Tournament will become a powerful engine for making meaningful progress on important protein engineering problems. Many of the most impactful protein engineering applications, from climate technology and green manufacturing to antivirals and diagnostics, are not being fully addressed by current research efforts. 

The Protein Engineering Tournament will become a vehicle for connecting cutting-edge researchers with the experimental resources necessary to make headway on important societal challenges that are not traditionally addressed by industry or academia (Figure 4A). While industry possesses significant resources to dedicate towards protein engineering problems, they are often restricted to applications which can recover the cost of research and generate profit. Conversely, while academia is capable of focusing on important problems without the constraints of profitability, research labs often lack the resources necessary to tackle large protein engineering efforts. This leaves a research gap where many applications with significant societal benefits have been under-resourced. With this consideration in mind, our protein engineering challenges will be selected with a strong focus on real-world impact.

The Tournament can be used to continually advance the field of protein engineering by selecting problems that push the limits of current techniques. Proteins possess a diverse array of functions, from enzymatic catalysis and molecular binding to chemical transportation, and our protein engineering challenges in the in silico and in vitro rounds will continuously evolve over time to reflect this myriad of functional possibilities. 

The first Tournament will likely focus on a single function, with enzyme engineering or protein binder design as strong initial candidates, but in future tournaments, the design challenges will expand to encompass more domains of function (Figure 4B). The order in which we introduce new functions will be driven by practical application, technical feasibility, and amenability to high-throughput experimentation. As our computational methods improve, our challenges will expand into increasingly more difficult and complex domains, such that the frontier of scientific capabilities is always represented in the Tournament’s challenges. 

The final protein engineering challenges will be decided on by our Target Selection team, with input from our scientific advisory board, philanthropic funding partners, and the larger scientific community. (Figure S1). The scientific advisory board will be composed of researchers in academia and industry. Their knowledge on the current frontier of challenges and opportunities in the field of protein engineering will further guide our selection of protein challenges. Additionally, we will invite input from funding partners who will use this opportunity to help advance the application of engineered proteins to problems salient to their philanthropic goals.

\section{TOURNAMENT CREATION ENGINE}
The Tournament represents a long-term investment in benchmarking the field of protein engineering. The tournament will produce new datasets for method development, new automated assays for protein characterization, and new benchmarked results for the community’s current-best algorithms, with the biennial cycle of the tournament also creating an opportunity to routinely take stock of current methods and assess the state of the field (Figure 5).

\begin{figure}
    \centering
    \includegraphics[width=1\columnwidth]{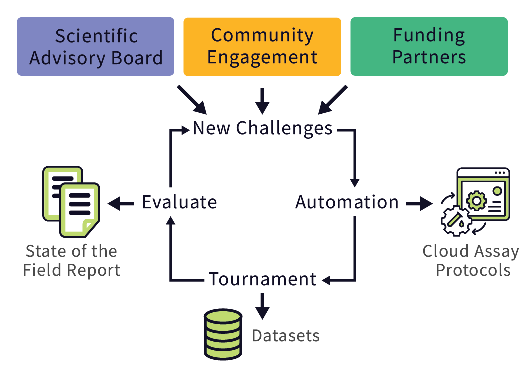}
    \caption{
        \textbf{Tournament Creation Engine.} Creating the Protein Engineering Tournaments will be a cyclical process. We will select new protein design challenges, develop automated assays for these challenges, host a tournament with these new assays and the datasets they’ve produced, and evaluate the tournament results. The assays, datasets, and evaluation will be open-sourced.}
    \label{fig:figure5}
\end{figure}

\subsection{Cloud laboratories and method development}
To maximize our impact, we want the assays we develop for each tournament to be available long after the tournament has concluded. To accomplish this goal and capitalize on the benefits of automated experimentation, we will design and execute our experimental workflows in cloud laboratories (Figure 5). 

Commercial cloud science laboratories, such as Emerald Cloud Labs, and academic cloud science laboratories, such as those found at Boston University and Carnegie Mellon, enable scientists to forgo the lab bench in favor of running experimental biology protocols in automation-enabled facilities. In this paradigm researchers write their assays in a symbolic laboratory programming language that specifies the instructions for each step of their protocol. Once written, scientists can queue up their experiments, wet-lab robots execute each step, and the resulting data is uploaded for analysis.

This approach offers the potential to greatly improve the accessibility and reproducibility of life science research by enabling scientists to share experimental protocols as easily as we share software. Experimental assays written with a symbolic lab language can be uploaded to code-sharing websites like Github, allowing any researcher from around the world to access and reproduce this work. Cloud laboratories will also improve the productivity of biological experimentation by transferring time-consuming experimental work from the hands of human researchers onto the decks of precise, high-throughput robots.

Cloud laboratories will serve as the experimental backbone of the Protein Engineering Tournament to ensure our in vitro experiments are high-throughput, reproducible, and accessible (Figure 5). 

At the conclusion of each Tournament, the protocols we developed to execute characterization assays in the in vitro round will be made openly available to the scientific community. Therefore, researchers will be able to continually 
benchmark their computational methods on the same standardized assays even after the Tournament has concluded. Since each Tournament will introduce new protein engineering challenges, this approach will lead to an ever-expanding corpus of open-source protein engineering workflows to help benchmark new computational methods for years to come. 

\subsection{Generating datasets}
The automated experimental protocols discussed above will be used to generate the datasets for the in silico and in vitro rounds and also be used to generate additional data based on participants’ submitted protein sequences during the in vitro round. After the Tournament has ended, all data produced during the Tournament will be made publicly available along with the corresponding automated protocols (Figure 5). Furthermore, the Tournament can act as an avenue for academic and corporate research entities to make unpublished datasets of protein function available as predictive challenges in the in silico round. 

\subsection{Analysis of the State of the Field}
Finally, we will aggregate the learnings from our tournament’s results into a State of the Field report (Figure 5). This report will analyze the performance of our participant’s computational approaches, noting the relative standing of different techniques across different challenges, and discussing our collective progress throughout the various domains of protein engineering.

\subsection{Governance}
The Protein Engineering Tournament is operated by Align to Innovate, a non-profit dedicated to improving the reproducibility, scalability, and shareability of life science research through community-driven initiatives. The Tournament will be run by a combination of Align to Innovate employees and volunteers from the protein engineering community (Figure S1). The Tournament Coordinator, who’s responsibilities entail coordinating the teams to ensure a successful tournament, will be a full time employee of Align to Innovate. Members of the Target Selection, Data Science, Organization and Outreach, and Cloud Assay Development teams will largely be composed of volunteers from the research community, in addition to supporting members from within the Align to Innovate team. 

\section{PILOT TOURNAMENT}
A pilot tournament began May 1st 2023 with the theme of Enzyme Design based on six datasets received from both industry and academic groups. Initial interest in the pilot tournament led to the registration of just over 30 teams, respresenting a mix of academic (55\%), industry (30\%), and independent (15\%) teams, with research experience running from Nobel Laureates to high school students. For the pilot tournament, the in vitro round experimentation will be performed in-house by a corporate partner. Development of the cloud laboratory assays for future tournaments is currently underway within Align to Innovate, the non-profit parent organization of the Protein Engineering Tournament. 

\section{CONCLUSION}
The Protein Engineering Tournament introduces an innovative, community-driven platform to accelerate the advancement of computational protein engineering. This open science initiative combines in silico and in vitro methods, employing cloud laboratories to ensure reproducibility and continued access to experimental workflows. By creating a unified benchmark, the Tournament stimulates collaboration and competition among researchers and presents opportunities for the community to test their computational models on novel protein engineering challenges. It offers a dynamic space for academia, industry, and independent groups to broaden the horizons of protein engineering, particularly in areas of high societal value that might be currently underserved. Further, through its integration with Align to Innovate, the Tournament builds upon the strengths of a diverse scientific community, fostering transparency and sharing of knowledge. As we look ahead to the first official Tournament, we anticipate that this initiative will contribute significantly to the evolving landscape of protein engineering, driving forward both technical and translational breakthroughs. By democratizing access to protein design, we aspire to inspire a new generation of computational protein engineers and guide the future of life science research.

\bibliographystyle{unsrtnat}
\bibliography{references}  






\end{document}